# Stabilization of helium shell burning by rotation in accreting white dwarfs

S.-C. Yoon and N. Langer

*Astronomical Institute, Utrecht University, Princetonplein 5, 3584 CC, Utrecht, The Netherlands*

**Abstract.** The currently favored scenario for the progenitor evolution of Type Ia supernovae (SNe Ia) presumes that white dwarfs in close binary systems grow to the Chandrasekhar limit via mass accretion from their non-degenerate companions. However, the accreted hydrogen and/or helium usually participate thermally unstable or even violent nuclear reactions in a geometrically confined region, due to the compactness of the white dwarf. Since shell flashes induced by the thermal instability might induce significant loss of mass, efficient mass increase of white dwarfs by hydrogen and/or helium accretion has been seriously questioned. A good understanding of the stability of thermonuclear shell sources is therefore crucial in order to investigate the evolution of accreting white dwarfs as SNe Ia progenitors. Here, we present a quantitative criterion for the thermal stability of thermonuclear shell sources, and discuss the effects of rotation on the stability of helium shell burning in helium accreting CO white dwarfs with $\dot{M} \approx 10^{-7}...10^{-6}$ $M_\odot$ yr$^{-1}$. In particular, we show that, if the effects of rotation are properly considered, helium shell sources are significantly stabilized, which might increase the likelihood for accreting white dwarfs to grow to the Chandrasekhar limit.

## INTRODUCTION

Many important astrophysical phenomena are related to the thermonuclear shell burning in accreting white dwarfs. Nova outbursts are induced by explosive hydrogen shell flash in highly degenerate hydrogen envelopes of white dwarfs which accretes hydrogen at a rate of $\dot{M} \approx 10^{-10} \ldots 10^{-8}$ $M_\odot$ yr$^{-1}$. Steady burning of hydrogen and/or helium in accreting white dwarfs with $\dot{M} \approx 10^{-7} \ldots 10^{-6}$ $M_\odot$ yr$^{-1}$ may be, on the other hand, relevant to the so-called super-soft X-ray sources (van den Heuvel et al. [2]). In particular, growth of accreting white dwarfs to the Chandrasekhar limit via such steady nuclear burning of accreted hydrogen and/or helium will lead to one of the most energetic events in the universe: Type Ia supernova (SN Ia).

Recent binary star evolution model by Yoon & Langer [7], who considered the evolution of a binary system consisting of a CO white dwarf and a helium giant, confirms that white dwarfs can reach the Chandrasekhar limit, at least, by helium accretion. However, accretion rates which allow steady nuclear shell burning in white dwarfs are limited to a very narrow range, for both hydrogen and helium. Further, steady burning of hydrogen usually leads to unstable helium shell flashes (Cassisi et al. [1]; Kato & Hachisu [3]). It is hence much debated whether hydrogen and/or helium accreting white dwarfs could be a major source for SNe Ia.

To have a better understanding on the issue, here we provide a quantitative criterion

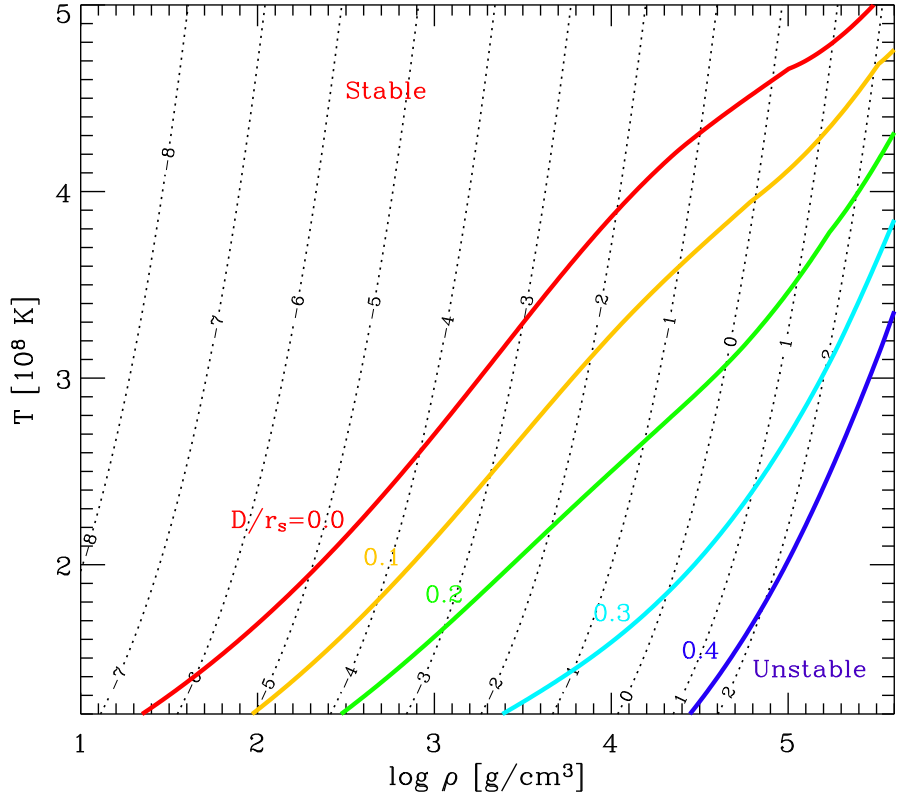

**FIGURE 1.** Stability conditions for a helium shell source in the density – temperature plane. The solid lines separate the thermally unstable region from the stable region, for 5 different relative shell source thicknesses (i.e., $D/r_s$ = 0.0, 0.1, 0.2, 0.3 and 0.4). The dotted contour lines denote the degeneracy parameter ($:= \psi/kT$). $X_{He} = 0.662$ and $X_C = 0.286$ have been assumed. From Yoon, Langer & van der Sluys [9]

for the stability of thermonuclear helium shell sources. We also present evolutionary models of helium accreting CO white dwarfs, where the effects of the centrifugal force on the white dwarf structure, and rotationally induced chemical mixing are taken into account.

## STABILITY OF THERMONUCLEAR SHELL SOURCES

Thermonuclear reaction rates are extremely sensitive to temperature and small increase in temperature may drastically enhance the energy generation. In the core of non-degenerate stars, however, increased energy output is consumed mostly for expansion work instead of increasing internal energy, and thermal stability is thus ensured (e.g. Kippenhahn & Weigert [4]).

In white dwarfs, on the other hand, accreted hydrogen and/or helium participate nuclear burning in a very dense and geometrically confined region in the white dwarf envelope. Since both electron degeneracy and strong geometrical confinement tend to keep local pressure in the shell source constant, nuclear reactions in the shell – especially in the helium burning shell – are prone to the thermal instability (cf. Schwarzschild & Härm [5]; Weigert [6]; Kippenhahn & Weigert [4]). In other words, a shell source becomes more susceptible to the thermal instability if it is more degenerate and geometrically thinner. Another factor to influence the stability of shell sources is temperature. With higher temperature, the role of radiation pressure becomes important, and sensitivity of nuclear reactions to a change in temperature becomes weaker, which favors thermal stability.

Yoon, Langer & van der Sluys [9] summarize conditions for the stability of helium shell sources, in terms of density, temperature, and relative thickness of the shell source, as in Fig. 1. In conclusion, the less dense, hotter, and thicker a shell source is, the more stable it becomes. See Yoon, Langer & van der Sluys [9] for more detailed discussion.

## STABILIZING EFFECTS OF ROTATION

White dwarfs accrete matter through the Keplerian disk, and believed to be spun up by the angular momentum gain from the accreted matter. Such a spin-up process might have two important effects on the behavior of helium shell burning in accreting white dwarfs.

Firstly, the centrifugal force will lift up the helium burning layers to reduce electron degeneracy, compared to the non-rotating case. Secondly, rotationally induced hydrodynamic instabilities in the spun-up layers will induce mixing of chemical elements. I.e., accreted helium will be mixed into the carbon-oxygen core, and helium burning layers will be accordingly extended. Both effects – decrease of degeneracy and widening of the shell source – favor thermal stability. Moreover, simulations of helium accreting white dwarfs by Yoon, Langer & Scheithauer [10], who considered the above mentioned effects, show that the helium shell source becomes somewhat hotter with rotation, mainly due to the increased $^{12}C(\alpha,\gamma)^{16}O$ reaction.

Fig. 2 clearly shows these stabilizing effects of rotation in helium accreting white dwarf models. In non-rotating models with $\dot{M} = 5 \times 10^{-7}$ $M_\odot$ yr$^{-1}$, thermal pulses induced by unstable helium shell burning derive the white dwarf envelope to reach the Eddington limit. In the corresponding rotating model, very weak thermal pulses appear soon after the mass accretion, but steady burning follows soon. With $\dot{M} = 10^{-6}$ $M_\odot$ yr$^{-1}$, helium shell burning remains stable even when the white dwarf grows to $\sim 1.5$ $M_\odot$. This result implies that mass accumulation of accreting white dwarfs might be much more efficient than believed before, in support of the currently favored scenario for the SNe Ia progenitor evolution (i.e., single degenerate Chandrasekhar mass scenario). Readers are referred to Yoon, Langer & Scheithauer [10] for more comprehensive discussion.

## CONCLUDING REMARKS

The above discussion indicates that rotation might play such a crucial role in the evolution of accreting white dwarfs that its effects should not be neglected in the study of SNe Ia progenitors. Apart from the stabilization of helium shell burning which may increase the mass accumulation efficiency of accreting white dwarfs, rotation may involve another important aspect. Differential rotation inside the white dwarf as a result of mass and angular momentum accretion might lead to SN Ia explosion at super-Chandrasekhar masses (Yoon & Langer [8]). Future work, including multi-dimensional studies on the effects of rotation (cf. Yoon & Langer 2005, in preparation), will reveal whether consideration of rotation would give better explanations of various aspects – such as production rate and diversity – of SNe Ia.

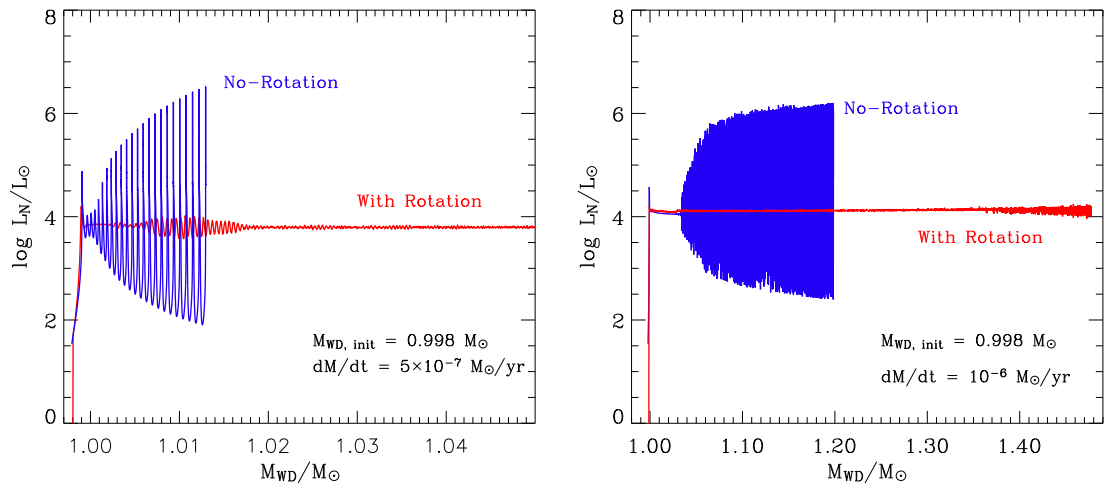

**FIGURE 2.** Evolution of the nuclear luminosity due to helium burning in helium accreting white dwarf models for $\dot{M} = 5 \times 10^{-7}\ M_\odot\ \mathrm{yr}^{-1}$ (left panel) and $\dot{M} = 10^{-6}\ M_\odot\ \mathrm{yr}^{-1}$ (right panel). The initial mass of the white dwarf is 0.998 $M_\odot$. The results with rotation and without rotation are given by the black and gray lines, respectively. From Yoon, Langer & Scheithauer [10].

# ACKNOWLEDGMENTS

This work has been supported by the Netherlands Organization for Scientific Research (NWO).